\documentclass[aps,prapplied,twocolumn,superscriptaddress,longbibliography]{revtex4-2}

\usepackage{amsmath}
\usepackage{amssymb}
\usepackage{graphicx}
\usepackage{dcolumn}
\usepackage{bm}
\usepackage{hyperref}
\usepackage{placeins}

\begin{document}

\title{Limits of Trap-assisted Photomultiplication Gain}

\author{Ardalan Armin}
\email{ardalan.armin@swansea.ac.uk}
\affiliation{Centre for Integrative Semiconductor Materials (CISM),
Department of Physics, Swansea University, Swansea SA1 8EN, United Kingdom}

\date{\today}

\begin{abstract}
Photodiodes based on trap-assisted current injection can exhibit internal
photomultiplication with apparent quantum efficiencies far exceeding unity,
raising the question of whether such gain fundamentally enhances detector
sensitivity. We employ a minimal analytical framework based on a single
gain-active trapped state coupling photogenerated carriers to contact injection.
The gain is intrinsically self-limiting: the injection process that amplifies the
current simultaneously accelerates relaxation of the gain-enabling state,
producing an inherently nonlinear, operating-point-dependent response. The form of
this nonlinearity is not universal---once the trap level is generalized to an
energetic distribution and recombination is allowed to be bimolecular, the same
mechanism yields superlinear, linear, or strongly sublinear responses. A single
chord gain is therefore not a meaningful device descriptor, and chord-gain
comparisons across the literature conflate devices in different regimes. Treating
trap occupancy and injection as coupled stochastic processes, we show that
internal gain introduces a strictly non-negative fluctuation penalty from the
dissipative dynamics that sustain the gain state. A local, small-signal
detectivity exhibits a finite optimum yet cannot exceed the intrinsic
thermodynamic limit of the underlying unity-gain photodiode. Gain is thus
equivalent to driven stochastic amplification: it can suppress downstream readout
noise, but cannot reduce the fundamental noise floor set by the primary
photodetection process.
\end{abstract}

\maketitle

\section{Introduction}

Photodetectors capable of registering weak optical signals are central to
applications ranging from biomedical imaging and environmental sensing to optical
communications and astronomy. In conventional photodiodes, the external quantum
efficiency (EQE) is bounded by unity since at most one electron is collected per
absorbed photon. Amongst a variety of photodetectors with internal gain
(EQE~$>1$), a distinct class of devices --- prominent in thin-film platforms based
on organic semiconductors, halide perovskites, nanocrystals, and hybrid systems
--- circumvents this bound through trap-assisted photomultiplication, a phenomenon
rooted in the barrier photoconductivity formulated by Petritz in the
1950s~\cite{Petritz1956} and first demonstrated in organic pigment films by
Hiramoto \textit{et al.}~\cite{Hiramoto1994}. In these devices, photogenerated
carriers become trapped near a contact, electrostatically modulating the injection
barrier for the opposite carrier type, thereby enabling amplified current flow
under reverse bias~\cite{Guo2012,Daanoune2020}. EQEs of $10^{2}$--$10^{5}$\% have
been widely reported across these material
systems~\cite{Li2015,Kublitski2021a,Wang2026}, though such values are typically
measured at low illumination intensities where the device operates in a strongly
nonlinear regime: such values represent a chord gain at a specific operating point, not an intrinsic material/device property. As we show, this distinction matters because the chord and tangential gains differ in a nonlinear device, and only the latter governs sensitivity.

This premise deserves careful scrutiny. The analogy most commonly invoked is with
photoconductivity, where gain arises from the ratio of carrier lifetime to transit
time as described by Petritz~\cite{Petritz1956} and formalised by
Rose~\cite{Rose1963}. In photoconductive detectors, it has long been understood
that the same mechanisms that enable high gain --- long carrier lifetimes and
trap-mediated recombination --- also enhance generation--recombination noise, so
that the input-referred sensitivity does not improve indefinitely with gain. In
trap-assisted photomultiplication photodiodes the situation is both analogous and
more complex because the gain arises often not from bulk transport but from the
electrostatic modulation of contact injection by trapped charge, making it
intrinsically coupled to interfacial energetics, injection kinetics, and the
applied bias. The operating point is therefore sustained by external electrical
work, placing the device in a nonequilibrium, dissipative regime fundamentally
different from that of a simple photodiode.

Despite the volume of experimental literature, as discussed in a recent
review~\cite{Wang2026}, several critical problems have gone largely unaddressed.
The first, and most consequential, is the systematic misreporting of specific
detectivity $D^{*}$ in gain-type photodetectors, as pointed out in previous
works~\cite{Fang2019,Bianconi2021,bianconi2021exaggerated}. The standard expression for $D^{*}$ assumes a
linear relationship between input optical power and output current so that an
extrapolation from the illuminated condition to the equilibrium condition is
possible. This assumption does not hold in photomultiplication photodiodes. The
noise then has contributions from stochastic trap kinetics and injection
fluctuations that are absent in the shot-noise formula~\cite{pelayo17}, and the
responsivity depends on the operating point set by the illumination level. As a
result, the $D^{*}$ values routinely extracted from photocurrent measurements at
finite illumination substantially overestimate the true
sensitivity~\cite{Rogalski}. Because both the noise and the responsivity are
intensity dependent, and because they must be evaluated at the same operating
point, a single global $D^{*}$ is not well-defined, and reported values --- often
measured at one illumination level and extrapolated to low-light conditions ---
can be physically misleading. Saggar \textit{et al.} have recently drawn attention
to this issue~\cite{Saggar2026},
reinforcing the need for a consistent theoretical framework.

The second problem concerns linearity and the linear dynamic range (LDR).
Photomultiplication photodiodes are inherently nonlinear. A conventional LDR is
defined as the range of optical input over which the output is linear; by this
definition, PM-type devices --- particularly in their gain regime --- have an LDR
that is either zero or poorly defined. As we show below, the nonlinearity is not
even of a single sign: depending on device construction and operating range, the
photocurrent can be superlinear, approximately linear, or sublinear over decades
of intensity, the latter being the case most frequently observed in practice. A
device may well be functional over a wide range of optical intensities, but this
should be described as a dynamic range (DR), not an LDR~\cite{Fang2019}. A direct
corollary is that the chord gain --- the ratio of total current to total photon
rate at a single intensity --- is not a transferable figure of merit, and the
common practice of comparing chord gains across the literature compares devices
sitting in qualitatively different regimes of the same underlying mechanism.

The third problem is conceptual and central to present work; whether
internal gain can, even in principle, improve the intrinsic detectivity of a
photodiode --- however that quantity is defined --- or whether it merely
redistributes signal and noise in the manner of a driven stochastic amplifier. A
related question concerns the role of dark current. In PM-type devices, the dark
operating point is set not only by thermally generated charges through
band-to-band or mid-gap trap transitions, but by the dark injection current
sustained by the applied bias, which is fundamentally different from the dark
current of a reverse-biased unity-gain photodiode~\cite{Sandberg2023,Kublitski2021b}
and is accompanied by its own noise budget.

The intention of this work is to address these issues within a minimal analytical
framework adapted from the classical theory of barrier
photoconductivity~\cite{Petritz1956,Carbone1994}, applied here specifically to
trap-assisted injection in photodiodes under reverse bias. The model is built on a
single gain-active trapped-charge state that couples photogenerated carriers to
contact injection, capturing the essential feedback mechanism underlying
photomultiplication while remaining independent of specific material systems or
device architectures. We deliberately keep this core model minimal: the
responsivity, noise, and detectivity results, and in particular the thermodynamic
bound, are properties of the operating point and do not depend on the microscopic
detail of how the trapped population is established. That detail does control the
shape of the current--intensity characteristic, and we introduce it where it
belongs --- in the analysis of linearity --- by generalizing the rate equation to
a distribution of traps and a bimolecular recombination channel, recovering the
single-level result as a limit. While the noise physics of barrier-modulated
photoconductors has been understood for
decades~\cite{Petritz1956,Carbone1994,Lummis1957,VanVliet1956}, and concerns about
the validity of reported figures of merit in PM photodiodes have been raised
recently~\cite{pelayo17,Fang2019,Bianconi2021,Saggar2026}, a unified treatment
connecting gain, noise, nonlinearity, and bandwidth in this device class ---
and drawing out the practical consequences for how sensitivity should be defined
and measured --- remains absent. This work provides that treatment.

\section{Photoresponsivity in the Presence of Gain}

Following the barrier photoconductivity framework of Petritz~\cite{Petritz1956}
and its subsequent development by
others~\cite{Carbone1994,Lummis1957,VanVliet1956}, we consider a photodiode under
reverse bias in which photogenerated carriers of one type become trapped near a
contact, thereby electrostatically modulating the injection barrier for the
counter carrier type. The gain mechanism operates through the following feedback
cycle. A photon is absorbed, generating an electron--hole pair. One carrier ---
say the electron --- becomes trapped near the cathode, while the hole is swept to
the anode by the applied field, contributing one elementary charge to the external
circuit as in a unity-gain photodiode. The trapped electron reduces the effective
injection barrier at the cathode, enabling hole injection from the positively
biased cathode into the active layer. Each injected hole transits the device and
is collected at the anode, contributing an additional elementary charge. The trap
is not necessarily relaxed by each injection event. If injected holes are
predominantly collected rather than recombined with the trapped electron, the trap
persists, the barrier remains lowered, and injection continues. The gain is
therefore the mean number of holes injected and collected before the trap is
finally relaxed, and many collection events arise from a single photogeneration
event. This places two physical requirements on the device: the trap must be
sufficiently long-lived that many injection events occur before thermal emission
relaxes it, and injected carriers must be predominantly collected rather than
recombine with the trap. The gain state is maintained by continuous electrical
work supplied by the applied bias, placing the device in a driven, nonequilibrium
regime, fundamentally distinct from a unity-gain photodiode operating close to
thermodynamic equilibrium.

To make this quantitative, we adopt a minimal model in which a single gain-active
trapped-charge state controls the injection current at the contact. The injected
current depends exponentially on the trap occupancy $n$ (the total number of
occupied traps) through the electrostatic modulation of the contact barrier. The
trapped charge reduces the effective barrier by $\alpha n$, where $\alpha$
quantifies the coupling between trap occupancy and the local interfacial
potential, so that $\phi = \phi_0 - \alpha n$. For a device of area $A$ with areal
trap density $\tilde{n} = n/A$ at distance $d$ from the contact, Poisson's equation
for an infinitely thin sheet of charge above a grounded conducting plane gives
$\alpha = qd/\varepsilon A$, where $\varepsilon = \varepsilon_r \varepsilon_0$ is
the permittivity of the active layer, $q$ the elementary charge. Defining the
dimensionless gain-control parameter
$\beta \equiv q\alpha/kT = q^{2}d/\varepsilon kTA$, with $k$ the Boltzmann constant
and $T$ the temperature, the injection current takes the form
\begin{equation}
  I(n) = I_0\, e^{\beta n},
  \label{eq:injection}
\end{equation}
where $I_0 \propto \exp(-q\phi_0/kT)$ is the dark injection current at the
intrinsic barrier $\phi_0$. Equation~(\ref{eq:injection}) is written in thermally
activated form but should be understood more generally as an effective description
of field-assisted injection --- including tunneling --- whose rate depends
exponentially on the interfacial electrostatic potential. This is an interfacial
effect, mechanistically distinct from classical photoconductivity in which gain
arises from the ratio of carrier lifetime to transit time~\cite{Rose1963}. The
parameter $\beta$ increases with trap proximity $d$ (traps closer to the contact
couple more strongly to the barrier) and decreases with permittivity, reflecting
dielectric screening of the trapped charge.

The trap occupancy evolves according to the balance between photoinduced filling,
thermal emission, and injection-assisted relaxation,
\begin{equation}
  \frac{dn}{dt} = \eta\Gamma - \frac{n}{\tau_t} - \frac{\gamma I(n)}{q},
  \label{eq:rate}
\end{equation}
where $\Gamma$ is the total absorbed photon rate, $\eta$ the internal quantum
efficiency for trap filling, $\tau_t$ the intrinsic trap lifetime, and $\gamma$
(dimensionless) the probability per injected carrier that the trap is relaxed
through recombination with the injected charge. The model operates in the regime
of small trap occupancy ($n \ll N_t$), where $N_t$ is the total trap density, so
that the trap-filling term is linear in the photon rate --- valid when the
quasi-Fermi level of the trapped carrier lies well below the trap energy, and the
regime relevant to weak-signal detection. The third term encodes the self-limiting
feedback of the gain: the same injection process that amplifies the current also
relaxes the gain-enabling state at rate $\gamma I/q$. For large gain, $\gamma$ must
be small --- most injected carriers must transit and be collected rather than
recombine with the trap --- while $\tau_t$ must be long enough that thermal
emission does not relax the trap before many injection events have occurred.

At steady state, $dn/dt = 0$, the trap occupancy $n^{*}$ and current
$I^{*} = I(n^{*})$ satisfy
\begin{equation}
  \eta\Gamma = \frac{n^{*}}{\tau_t} + \frac{\gamma I^{*}}{q}.
  \label{eq:steadystate}
\end{equation}
This steady state is intrinsically nonequilibrium; it is sustained by continuous
optical generation and bias-driven injection. The right-hand side represents two
dissipative relaxation channels --- thermal emission at rate $n^{*}/\tau_t$ and
injection-assisted relaxation at rate $\gamma I^{*}/q$ --- both maintained by
external energy input via reverse bias. The electrical power $I^{*}V$ supplied by
the bias is continuously dissipated to maintain this state. Gain therefore comes
at an irreducible thermodynamic cost, and any noise analysis must account for the
stochasticity of both dissipative channels.

To determine the responsivity, we linearize Eq.~(\ref{eq:rate}) about the steady
state $(n^{*}, I^{*})$. Writing $n = n^{*} + \delta n$ and using
\begin{equation}
  \delta I = \left.\frac{\partial I}{\partial n}\right|_{n^{*}}\!\delta n
           = \beta I^{*}\delta n,
  \label{eq:linearI}
\end{equation}
and subtracting Eq.~(\ref{eq:steadystate}), the linearized equation for the
occupancy perturbation is
\begin{equation}
  \frac{d\,\delta n}{dt} = \eta\,\delta\Gamma - \lambda\,\delta n,
  \label{eq:linearized}
\end{equation}
where the effective relaxation rate is
\begin{equation}
  \lambda = \frac{1}{\tau_t} + \frac{\gamma\beta I^{*}}{q}.
  \label{eq:lambda}
\end{equation}
The rate $\lambda$ governs how quickly a perturbation in trap occupancy decays. It
has two additive contributions: the intrinsic thermal emission rate $1/\tau_t$, and
the injection-assisted relaxation rate $\gamma\beta I^{*}/q$, which grows with the
operating-point current and is therefore intensity dependent --- a dependence that
underlies the nonlinearity, intensity-dependent bandwidth, and
operating-point-dependent noise derived below.

Taking the Fourier transform of Eq.~(\ref{eq:linearized}) and using
$\delta I = \beta I^{*}\delta n$, the small-signal current perturbation is
\begin{equation}
  \delta I(\omega) = \frac{\beta I^{*}\eta}{\lambda + i\omega}\,\delta\Gamma(\omega).
  \label{eq:deltaI}
\end{equation}
The optical responsivity is $R = \delta I/\delta P$, where
$\delta P = h\nu\,\delta\Gamma$ is the optical power perturbation, with $h\nu$ the
photon energy, giving
\begin{equation}
  R(\omega) = \frac{\eta\beta I^{*}}{h\nu(\lambda + i\omega)}.
  \label{eq:R_omega}
\end{equation}
In the low-frequency limit this reduces to
\begin{equation}
  R(0) = \frac{\eta q}{h\nu}\,G_\mathrm{eff},
  \label{eq:R0}
\end{equation}
where the effective gain is
\begin{equation}
  G_\mathrm{eff} = \frac{\beta I^{*}}{q\lambda}
                 = \frac{\beta I^{*}/q}{1/\tau_t + \gamma\beta I^{*}/q}.
  \label{eq:Geff}
\end{equation}
$G_\mathrm{eff}$ is the mean number of carriers injected and collected per trap
filling event before relaxation. For $G_\mathrm{eff} = 1$, Eq.~(\ref{eq:R0})
reduces to the standard result $R = \eta q/h\nu$, recovering the
quantum-efficiency-limited responsivity of a unity-gain photodiode.

Equations~(\ref{eq:R0}) and~(\ref{eq:Geff}) expose the self-limiting character of
the gain through its thermodynamic structure. The numerator $\beta I^{*}/q$ is the
rate at which the injection channel responds to the trap occupancy --- a rate that
grows with $I^{*}$ and hence with the power dissipated at the contact. The
denominator $\lambda$ is the total relaxation rate of the trapped state, which also
grows with $I^{*}$ through the injection-assisted term $\gamma\beta I^{*}/q$.
Because both the signal sensitivity and the relaxation rate are driven by the same
kinetics, they cannot be independently optimized: increasing $I^{*}$ to raise the
gain simultaneously accelerates the decay of the gain-active state. More
fundamentally, the energy required to sustain $I^{*}$ is supplied by the bias and
dissipated through the very channel that enables amplification. The gain is
therefore not a reversible signal multiplier but a dissipative amplification
process, and the fluctuations associated with that dissipation constitute an
irreducible noise penalty derived below.

\section{Linearity, and the Inadequacy of Chord Gain}

The minimal model of the preceding section makes a definite prediction for the
current--intensity characteristic. At low intensity the injection-assisted
relaxation term $\gamma I^{*}/q$ is negligible relative to the thermal emission
rate $1/\tau_t$, and Eq.~(\ref{eq:steadystate}) reduces to
$n^{*} \approx \eta\Gamma\tau_t$. The current then grows as
$I^{*} \approx I_0\exp(\beta\eta\Gamma\tau_t)$, which is strongly superlinear in
$\Gamma$ owing to the exponential sensitivity of injection to trap occupancy. As
$\Gamma$ increases, $\gamma I^{*}/q$ becomes significant, the steady-state
condition approaches $\eta\Gamma \approx \gamma I^{*}/q$, the current becomes
proportional to optical intensity, and the effective gain saturates toward
$G_\mathrm{eff} \to 1/\gamma$. The crossover between these regimes is governed by
$\gamma\beta I^{*}/q \sim 1/\tau_t$. The minimal model therefore predicts a
superlinear-to-linear response.

This prediction rests on two idealizations that are rarely met. The first is a
single, sharp trap level whose occupancy is limited only by its own lifetime; the
second is a barrier lowering, and hence an injection enhancement, that grows
without bound with occupancy through the unsaturated exponential of
Eq.~(\ref{eq:injection}). Experimentally, the opposite behaviour is the norm so that the
measured photocurrent is sublinear over a wide range of intensities, frequently
following a power law $I^{*}\!-\!I_0 \propto \Gamma^{\,s}$ with an exponent $s$
near one-half~\cite{Liu2023,Guo2020}. To capture both behaviours, and the continuum
between them, within a single framework, we relax these idealizations while
retaining the gain mechanism of Eq.~(\ref{eq:injection}).

We replace the single level by an exponential distribution of trap states below
the transport edge,
\begin{equation}
  g(\varepsilon) = \frac{N_t}{kT_c}\,e^{-\varepsilon/kT_c},
  \label{eq:dos}
\end{equation}
where $\varepsilon\ge 0$ is the energy depth below the band edge and $T_c>T$ is the
characteristic temperature of the distribution, a standard description of energetic
disorder in these material
systems~\cite{Schmidlin1977,Noolandi1977,TiedjeRose1981}. Under steady
illumination the carrier population partitions, in multiple-trapping
quasi-equilibrium, between mobile states of density $n_\mathrm{free}$ at the
transport edge and trapped states of total occupancy $n_t$. With an electron
quasi-Fermi level at depth $\varepsilon_F$, the mobile and trapped densities are
$n_\mathrm{free}=N_c\,e^{-\varepsilon_F/kT}$ and
$n_t \approx N_t\,e^{-\varepsilon_F/kT_c}$, where $N_c$ is the effective density of
transport states, so that
\begin{equation}
  n_t = N_t\!\left(\frac{n_\mathrm{free}}{N_c}\right)^{\!T/T_c}.
  \label{eq:nt_nfree}
\end{equation}
The trapped charge sets the gain. It is the trapped sheet that electrostatically
lowers the injection barrier, so the current depends exponentially on $n_t$,
$I^{*}=I_0\exp(\beta n_t)$, as in Eq.~(\ref{eq:injection}), now with $n_t$ given by
Eq.~(\ref{eq:nt_nfree}). The
steady-state generation balance is set by the loss of the mobile carriers, which
recombine to the ground state through a monomolecular channel of lifetime
$\tau_r$ and a bimolecular channel, together with the injection-assisted
relaxation of Eq.~(\ref{eq:rate}),
\begin{equation}
  \eta\Gamma = \frac{n_\mathrm{free}}{\tau_r}
             + \zeta\gamma_L\,n_\mathrm{free}^{2}
             + \frac{\gamma I^{*}}{q}.
  \label{eq:balance_gen}
\end{equation}
Here $\gamma_L = q(\mu_n+\mu_p)/\varepsilon$ is the Langevin recombination
coefficient~\cite{Langevin1903}, with $\mu_n,\mu_p$ the carrier mobilities, and
$\zeta\le 1$ is the Langevin reduction factor that quantifies the suppression of
bimolecular recombination below the Langevin rate, primarily through redissociation of charge-transfer states into free charges ~\cite{Armin2018,Braun1984}. We have taken the
mobile electron and hole densities to be comparable in the recombination zone, so
that the bimolecular rate scales as $n_\mathrm{free}^{2}$. The single discrete
level of Sec.~II is recovered in the limit of vanishing energetic disorder, for
which $n_t \propto n_\mathrm{free}$ (unit exponent in Eq.~(\ref{eq:nt_nfree})) and
$n_t \propto \Gamma$; Eqs.~(\ref{eq:nt_nfree}) and~(\ref{eq:balance_gen}) then
reduce to Eq.~(\ref{eq:steadystate}).

Three regimes follow directly. At the lowest intensities the monomolecular channel
dominates the loss, $\eta\Gamma \approx n_\mathrm{free}/\tau_r$, so
$n_\mathrm{free}\propto\Gamma$ and, through Eq.~(\ref{eq:nt_nfree}),
\begin{equation}
  n_t \propto \Gamma^{\,T/T_c}.
  \label{eq:mono_exp}
\end{equation}
For weak coupling, $\beta n_t \ll 1$, the injection is effectively linearized,
$\exp(\beta n_t)\approx 1+\beta n_t$, so the photocurrent simply tracks the trapped
charge, $I^{*}-I_0 \approx I_0\beta n_t \propto \Gamma^{\,T/T_c}$, a sublinear power
law whose exponent is set entirely by the energetic disorder. At higher
intensities the reduced-Langevin channel takes over, $\eta\Gamma \approx
\zeta\gamma_L n_\mathrm{free}^{2}$, so $n_\mathrm{free}\propto\Gamma^{1/2}$ and the
trapped-charge exponent halves,
\begin{equation}
  n_t \propto \Gamma^{\,T/2T_c},
  \label{eq:bi_exp}
\end{equation}
producing a still more sublinear photocurrent. The intensity at which the response transitions from Eq.~(\ref{eq:mono_exp}) to Eq.~(\ref{eq:bi_exp}) is fixed by the
mobile density $n_\mathrm{crit}=1/(\zeta\gamma_L\tau_r)$ at which the two
recombination channels balance, and hence moves with the reduction factor $\zeta$:
weaker bimolecular recombination (smaller $\zeta$) pushes the crossover to higher
intensity and extends the monomolecular regime.

The barrier coupling $\beta n_t$ --- the barrier lowering in units of $kT$ --- is
the control parameter for linearity. While $\beta n_t \lesssim 1$ the injection stays in its linear regime, so the
photocurrent remains proportional to the trapped charge and follows the sublinear
power laws of Eqs.~(\ref{eq:mono_exp}) and~(\ref{eq:bi_exp}). Once $\beta n_t
\gtrsim 1$, the exponential injection reasserts itself. Because $n_t$ continues to
grow with intensity, $\exp(\beta n_t)$ drives the photocurrent superlinear, and the
onset moves to lower intensity as $\beta = q^{2}d/\varepsilon kTA$ increases ---
that is, for traps closer to the contact, lower permittivity, or lower
temperature. The gain mechanism thus does not disappear under disorder; it is
delayed until enough trapped charge has accumulated, and whether it ever dominates
the measured range depends on the device construction.

\begin{figure*}[t]
  \centering
  \includegraphics[width=\textwidth]{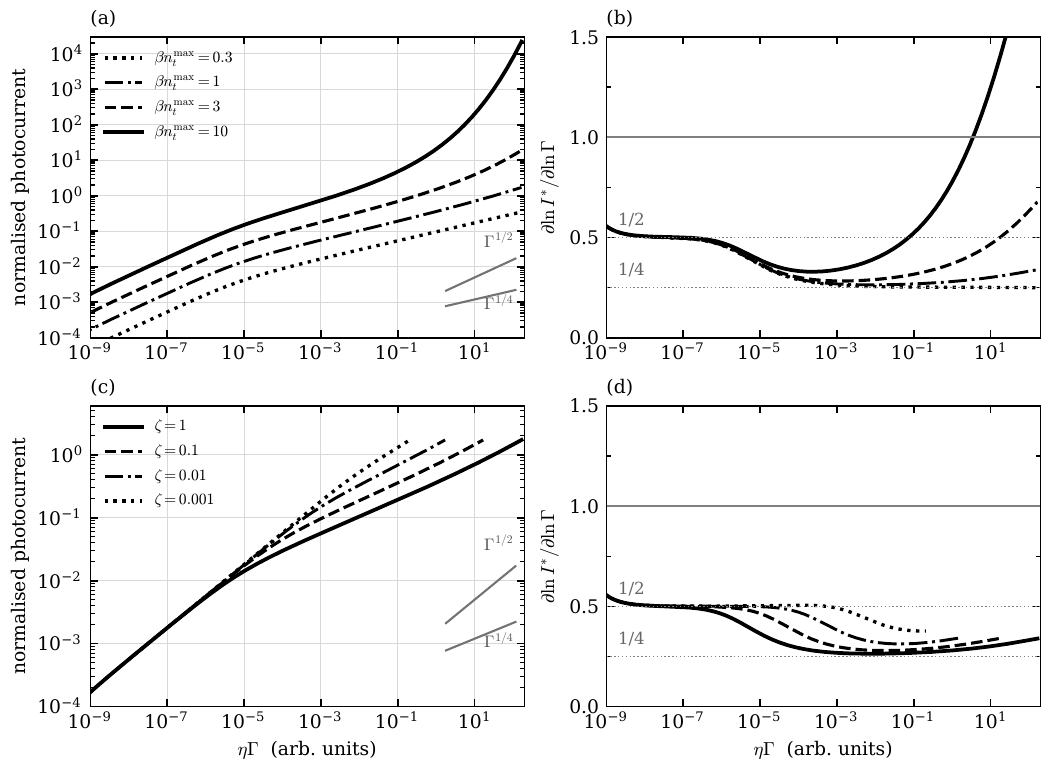}
  \caption{\textbf{Nonlinearities in trap-assisted photodetectors} Normalised
    photocurrent and its local exponent $\partial\ln I^{*}/\partial\ln\Gamma$
    versus absorbed photon rate, from the self-consistent solution of
    Eqs.~(\ref{eq:nt_nfree}) and~(\ref{eq:balance_gen}) with $T/T_c=1/2$. The
    normalised photocurrent is the photo-induced current $I^{*}-I_0$ referenced to
    the dark injection current $I_0$, i.e.\ $(I^{*}-I_0)/I_0$.
    (a,b)~Coupling sweep at full Langevin recombination ($\zeta=1$) for
    $\beta n_t^{\max}=0.3,1,3,10$. All curves share the low-intensity exponent
    $T/T_c=1/2$ (and tighten toward $T/2T_c=1/4$ where bimolecular recombination
    dominates); strong coupling drives the exponent through unity into
    superlinearity at high intensity. (c,d)~Reduction sweep at fixed
    $\beta n_t^{\max}=1$ for $\zeta=1,0.1,0.01,0.001$. The reduction factor only
    shifts the $\Gamma^{1/2}\!\to\Gamma^{1/4}$ crossover along the intensity axis;
    the exponents are fixed by the disorder and the recombination order. Guides of
    slope $1/2$ and $1/4$ are shown in gray.}
  \label{fig:linearity}
\end{figure*}

These behaviours are summarized in Fig.~\ref{fig:linearity}, obtained by solving
Eqs.~(\ref{eq:nt_nfree}) and~(\ref{eq:balance_gen}) self-consistently with
$T/T_c=1/2$. Panels (a) and (b) sweep the coupling at full Langevin recombination
($\zeta=1$): every curve shares the same low-intensity exponent $T/T_c$, because
$\beta n_t\to 0$ as $\Gamma\to 0$ for all of them, and the curves fan out only at
higher intensity, where strong coupling drives the local exponent up through unity
into superlinearity. Panels (c) and (d) fix the coupling at $\beta n_t^{\max}=1$
and sweep the reduction factor: the only effect of $\zeta$ is to slide the
monomolecular-to-bimolecular crossover --- the transition from exponent $T/T_c$ to
$T/2T_c$ --- along the intensity axis, leaving the exponents themselves unchanged.

These systems are intrinsically nonlinear, and the form of
the nonlinearity is not universal. The sublinear exponents are set by the energetic
disorder ($T/T_c$) and the order of the dominant recombination (the factor of two
between monomolecular and bimolecular control). The Langevin reduction factor and
the barrier coupling $\beta$ --- itself fixed by the trap--contact separation, the
permittivity, and the trap density --- set only the intensities at which the
device crosses between regimes and whether the exponential injection ever drives
the response superlinear. Consequently the same mechanism, and even the same
material, can present as superlinear, linear, or sublinear depending on operating
range and device construction. A single chord gain --- the ratio of total current
to total photon rate at one intensity --- is therefore not a meaningful device
descriptor, and the widespread practice of comparing chord gains across the
literature conflates devices operating in different regimes. The physically
meaningful quantity is the tangential (differential) gain at a specified operating
point, $\delta I/h\nu\,\delta\Gamma$, which is what enters the responsivity
Eq.~(\ref{eq:R_omega}) and the detectivity analysis below.

Because the $I$-vs-$\Gamma$ characteristic is nonlinear, the chord gain
$I^{*}/h\nu\Gamma$ and the tangential gain $\delta I/h\nu\delta\Gamma$ do not in
general coincide, and the discrepancy is operating-point dependent. By
construction, the LDR in the strict sense is therefore either zero or confined to
a narrow window. What is often reported as LDR in the literature is more properly a
dynamic range, the ratio of largest to smallest detectable signal --- a meaningful
but distinct quantity that does not imply linearity. Conflating the two
misrepresents the operating characteristics of these devices and sets expectations
that cannot be met in applications requiring a linear input--output relationship.

The same intensity dependence shapes the bandwidth. The frequency-dependent
responsivity in Eq.~(\ref{eq:R_omega}) is a first-order low-pass response with
$3$~dB cutoff
\begin{equation}
  \omega_{3\mathrm{dB}} = \lambda = \frac{1}{\tau_t} + \frac{\gamma\beta I^{*}}{q}.
  \label{eq:bw}
\end{equation}
Since $I^{*}$ increases with $\Gamma$, the bandwidth is not a fixed device
characteristic but grows with optical intensity. At low intensity it is limited by
the intrinsic trap lifetime, $\omega_{3\mathrm{dB}} \approx 1/\tau_t$, consistent
with the slow response commonly observed in high-gain PM photodiodes; at high
intensity, injection-assisted relaxation dominates and
$\omega_{3\mathrm{dB}} \approx \gamma\beta I^{*}/q \propto \Gamma$. The device
becomes faster precisely as the gain saturates --- a direct consequence of the
shared dynamics. (In a disordered device the single corner frequency of
Eq.~(\ref{eq:bw}) is dispersed into a distribution of relaxation rates, with
consequences for the noise spectrum that we address in Sec.~IV.) Responsivity and
bandwidth should therefore always be reported at the same operating intensity,
since neither is a fixed device constant and their product is governed by the
dissipative injection dynamics of the gain state.

\section{Noise}

The gain state $(n^{*}, I^{*})$ is sustained by three elementary stochastic
processes: photoinduced trap filling at rate $\eta\Gamma$, thermal emission at rate
$n^{*}/\tau_t$, and injection-assisted relaxation at rate $\gamma I^{*}/q$. Each is
a Poisson process driven either by the optical field or by the bias-sustained
carrier flux, and each carries an independent generation--recombination (GR) noise
contribution. This is a direct consequence of the nonequilibrium, dissipative
nature of the operating point: a system maintained by continuous carrier flux
cannot avoid the fluctuations intrinsic to that flux. We treat these fluctuations
with a Langevin approach, introducing independent noise sources $\xi_i$ associated
with each elementary process. The linearized stochastic equation for the trap
occupancy is
\begin{equation}
  \frac{d\,\delta n}{dt} = -\lambda\,\delta n + \eta\,\delta\Gamma
    + \xi_\mathrm{cap} - \xi_\mathrm{em} - \xi_\mathrm{rel},
  \label{eq:langevin}
\end{equation}
where the low-frequency noise strengths are Poissonian, $S_{\xi_i} = 2R_i$, with
$R_i$ the corresponding mean transition rate. At steady state
\begin{equation}
  R_\mathrm{cap}^{*} = \eta\Gamma,\quad
  R_\mathrm{em}^{*} = \frac{n^{*}}{\tau_t},\quad
  R_\mathrm{rel}^{*} = \frac{\gamma I^{*}}{q}.
  \label{eq:rates}
\end{equation}
Setting $\delta\Gamma = 0$ and solving in Fourier space, the power spectral density
(PSD) of trap occupancy fluctuations is
\begin{equation}
  S_n(\omega) = \frac{2\eta\Gamma + 2n^{*}/\tau_t + 2\gamma I^{*}/q}{\lambda^2 + \omega^2},
  \label{eq:Sn}
\end{equation}
a Lorentzian with corner frequency $\lambda$, as expected for a single fluctuating
two-state system. The numerator is the sum of all three transition rates, each
appearing with weight~$2$ as required by the fluctuation--dissipation structure of
a driven system. Crucially, the injection-assisted relaxation term
$2\gamma I^{*}/q$ is absent in a unity-gain photodiode --- it is noise generated
entirely by the dissipative process that sustains the gain state, and cannot be
eliminated without eliminating the gain itself.

Since $\delta I = \beta I^{*}\delta n$, the trap-fluctuation current noise is
\begin{equation}
  S_{I,\mathrm{tr}}(\omega) = (\beta I^{*})^2 S_n(\omega)
    = \frac{(\beta I^{*})^2\!\left(2\eta\Gamma + 2n^{*}/\tau_t
      + 2\gamma I^{*}/q\right)}{\lambda^2 + \omega^2}.
  \label{eq:SItr}
\end{equation}
In addition, the injection current contributes shot noise independently of the
trap fluctuations,
\begin{equation}
  S_{I,\mathrm{inj}} = 2qI^{*}.
  \label{eq:SIinj}
\end{equation}
The underlying unity-gain photodiode contributes an intrinsic noise floor $S_0$,
encompassing the shot noise of the dark generation current and any Johnson noise of
the device resistance. The total current noise is therefore
\begin{equation}
  S_I(\omega) = \frac{(\beta I^{*})^2\!\left(2\eta\Gamma + 2n^{*}/\tau_t
    + 2\gamma I^{*}/q\right)}{\lambda^2 + \omega^2} + 2qI^{*} + S_0.
  \label{eq:SItotal}
\end{equation}
All contributions depend on the steady-state operating point through
$I^{*}(\Gamma)$ and $n^{*}(\Gamma)$ and are therefore intrinsically intensity
dependent --- a defining feature of trap-assisted photodiodes that must be
accounted for whenever noise is used to infer sensitivity.

To connect the noise to sensitivity, we refer all contributions back to the optical
input via the small-signal responsivity $R(\omega)$ from Eq.~(\ref{eq:R_omega}).
Because the device is nonlinear, both the responsivity and the noise must be
evaluated at the same operating point $\Gamma$. A single global noise-equivalent
power is therefore not well defined; the appropriate quantity is a local,
small-signal NEP defined for perturbations about the steady state,
\begin{equation}
  \mathrm{NEP}_\mathrm{loc}^2(\Gamma,\omega) = \frac{S_I(\omega)}{|R(\omega)|^2}
    = \mathrm{NEP}_\mathrm{tr}^2 + \mathrm{NEP}_\mathrm{inj}^2(\omega)
      + \mathrm{NEP}_0^2.
  \label{eq:NEPloc}
\end{equation}
Substituting Eqs.~(\ref{eq:R_omega}) and~(\ref{eq:SItr}), the $\omega$-dependent
Lorentzian factors cancel exactly between $S_{I,\mathrm{tr}}(\omega)$ and
$|R(\omega)|^2$, since both share the relaxation rate $\lambda$ --- the trap
fluctuations and the signal response are governed by the same dynamics. This
cancellation makes $\mathrm{NEP}_\mathrm{tr}^2$ frequency independent. The
trap-fluctuation contribution is
\begin{equation}
  \mathrm{NEP}_\mathrm{tr}^2
    = \left(\frac{h\nu}{\eta}\right)^{\!2}
      \!\left(2\eta\Gamma + \frac{2n^{*}}{\tau_t} + \frac{2\gamma I^{*}}{q}\right)
    = 4\eta\Gamma\!\left(\frac{h\nu}{\eta}\right)^{\!2}.
  \label{eq:NEPtr}
\end{equation}
The three terms are independent GR contributions from photogeneration, thermal
emission, and injection-assisted relaxation. Since each is a Poisson process with
noise power $2R_i$, and the steady-state constraint requires
$R_\mathrm{em}^{*} + R_\mathrm{rel}^{*} = \eta\Gamma$, the total is
$2\eta\Gamma + 2\eta\Gamma = 4\eta\Gamma$, giving the compact form in
Eq.~(\ref{eq:NEPtr}). This is directly analogous to the GR noise factor of~$2$
identified by Rose~\cite{Rose1963} in photoconductive systems --- each
photogeneration event creates noise both at the filling step and at the relaxation
step.

The injection shot-noise contribution, in contrast, is gain-dependent. In the
low-frequency limit
\begin{equation}
  \mathrm{NEP}_\mathrm{inj}^2(0)
    = \frac{2qI^{*}}{|R(0)|^2}
    = \left(\frac{h\nu}{\eta}\right)^{\!2}\frac{2q\lambda^2}{\beta^2 I^{*}},
  \label{eq:NEPinj0}
\end{equation}
and at finite frequency
\begin{equation}
  \mathrm{NEP}_\mathrm{inj}^2(\omega)
    = \left(\frac{h\nu}{\eta}\right)^{\!2}
      \frac{2q(\lambda^2 + \omega^2)}{\beta^2 I^{*}},
  \label{eq:NEPinjw}
\end{equation}
which grows with frequency since $S_{I,\mathrm{inj}}$ is white while
$|R(\omega)|^2$ rolls off as a Lorentzian. The low-frequency injection shot noise
decreases with increasing $I^{*}$, reflecting the growing transconductance of the
injection channel, and is maximal at the dark operating point $I^{*}\to I_0$.

The total low-frequency local NEP is
\begin{equation}
  \mathrm{NEP}_\mathrm{loc}^2(\Gamma)
    = \mathrm{NEP}_0^2
    + \left(\frac{h\nu}{\eta}\right)^{\!2}\!
      \left[4\eta\Gamma + \frac{2q\lambda^2}{\beta^2 I^{*}}\right].
  \label{eq:NEPtotal}
\end{equation}
This is the central result of the noise analysis. The first bracketed term is the
total trap GR noise; the second, $2q\lambda^2/\beta^2 I^{*}$, is the input-referred
injection shot noise --- the sole gain-specific noise penalty not determined by the
generation rate alone. It decreases with $I^{*}$ while the photogeneration shot
noise increases, producing a finite minimum in the total NEP as a function of
operating point. This competition and its consequences for detectivity are the
subject of the following section.

We introduce the dimensionless gain-state parameter
$x \equiv \gamma\beta I^{*}\tau_t/q$ --- the ratio of injection-assisted to thermal
relaxation rates --- and the dimensionless photon flux
$p \equiv \eta\Gamma\tau_t$. In terms of $x$, the steady-state current and
relaxation rate are $I^{*} = xq/\gamma\beta\tau_t$ and $\lambda = (1+x)/\tau_t$. We
normalize all noise contributions by the characteristic scale
\begin{equation}
  \mathrm{NEP}_\star^2 = \left(\frac{h\nu}{\eta}\right)^{\!2}\frac{2}{\tau_t},
  \label{eq:NEPstar}
\end{equation}
the input-referred noise power of the trap system set by the intrinsic thermal
emission rate $1/\tau_t$ alone --- the fundamental noise scale of the trap
independent of any gain mechanism. Normalizing Eq.~(\ref{eq:NEPtotal}),
\begin{equation}
  \tilde{N}(x,p) = \tilde{N}_0 + 2p + r\frac{(1+x)^2}{x},
  \label{eq:Ntilde}
\end{equation}
where $\tilde{N}_0 = \mathrm{NEP}_0^2/\mathrm{NEP}_\star^2$ and
$r \equiv \gamma/\beta$ (dimensionless). The term $2p$ contains the full trap GR
noise --- photon shot noise plus both relaxation-channel noise powers. The term
$r(1+x)^2/x$ is the normalized injection shot noise, the only term that depends
explicitly on the gain-state parameter $x$. As first noted by Petritz for
barrier-modulated photoconductors~\cite{Petritz1956}, ``barrier amplification does
not affect $S_s(G{-}R)$ because both signal and noise are amplified''; the present
analysis generalises this result to the dissipative injection case, where an
additional gain-specific noise penalty appears.

The preceding analysis assumes a single trap relaxation rate $\lambda$, yielding a
Lorentzian spectrum, Eq.~(\ref{eq:Sn}). A realistic device has traps distributed in
energy, as invoked in Sec.~III, and each depth $\varepsilon$ contributes a
relaxation time $\tau(\varepsilon)=\nu^{-1}e^{\varepsilon/kT}$ ($\nu$ an attempt
frequency); the occupancy spectrum is then a superposition of Lorentzians. An
exponential trap distribution maps onto an approximately uniform distribution of
$\ln\tau$, which is the classical McWhorter route to a $1/f$-type
spectrum~\cite{McWhorter1957,DuttaHorn1981}, so that the white GR floor derived
above is in practice coloured toward $1/f$ over the band bounded by the fastest and
slowest trap rates. Contact injection carries its own number- and
mobility-fluctuation $1/f$ noise~\cite{Hooge1994} that lies outside the present
treatment and is frequently the dominant low-frequency floor in real devices. None
of this alters the input-referred results that matter here. The cancellation
between $S_{I,\mathrm{tr}}(\omega)$ and $|R(\omega)|^2$ is exact Lorentzian by
Lorentzian, because signal and trap fluctuations are read out through the identical
kinetics, Eqs.~(\ref{eq:R_omega}) and~(\ref{eq:SItr}); it therefore survives any
superposition of relaxation times, and the input-referred GR factor
$4\eta\Gamma(h\nu/\eta)^2$, Eq.~(\ref{eq:NEPtr}) --- with it the detectivity bound
of the next section --- is independent of the spectral shape. A distribution of
timescales changes the frequency at which a given $D^{*}_\mathrm{loc}$ is realised,
not its value or its bound; $D^{*}_\mathrm{loc}$ must accordingly be quoted at a
specified frequency, and injection $1/f$ noise stands as an additional,
device-specific floor. The minimal single-trap model is thus sufficient for
establishing the input-referred noise and the thermodynamic limit that are the
object of this work, and we retain it for the remainder of the analysis.

\section{Specific Detectivity}

The specific detectivity $D^{*}$ is the standard figure of merit for photodetector
sensitivity, normalizing the signal-to-noise ratio per unit optical input, unit
bandwidth, and unit detector area. Its conventional definition assumes a linear
input--output relationship and an intensity-independent noise --- neither of which
holds in trap-assisted PM photodiodes. The responsivity depends on the operating
point through $I^{*}(\Gamma)$, the noise depends on $\Gamma$ through all
dissipative channels as established in Eq.~(\ref{eq:SItotal}), and the device is
inherently nonlinear with an operating-point-dependent exponent as shown in
Sec.~III. A single global $D^{*}$ is therefore not well defined for this class of
devices.

\begin{figure}[t]
  \centering
  \includegraphics[width=\columnwidth]{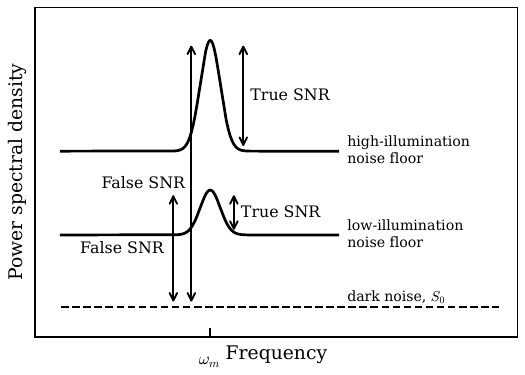}
  \caption{\textbf{True versus false signal-to-noise ratio.} Schematic power
  spectral density of the output current at two DC illumination levels,
  $\Gamma_1 < \Gamma_2$, modulated at $\omega_m$ with $\delta\Gamma \ll \Gamma$.
  The true SNR references the signal peak to the noise floor present at that
  operating point; the false SNR references it to the dark floor $S_0$ (dashed).
  In a photomultiplication photodiode the illuminated floor lies well above
  $S_0$, not merely by the photon shot noise common to any photodiode but
  through gain-specific terms --- the injection shot noise $2qI^{*}$ and the
  generation--recombination noise of the dissipative relaxation channels,
  Eq.~(\ref{eq:SItotal}) --- which grow with $I^{*}(\Gamma)$. Referencing to
  $S_0$ thus overestimates the sensitivity, increasingly so at higher
  illumination. A consistent $D^{*}_\mathrm{loc}$ requires signal and noise from
  the same spectrum at the same operating point.}
\label{fig:PSD}
\end{figure}

Reporting a $D^{*}$ extracted from a photocurrent measurement at finite
illumination using the shot-noise formula --- or from noise measured in the dark,
thereby ignoring the gain-state dependence of the noise floor in
Eq.~(\ref{eq:SItotal}) --- conflates chord gain with tangential gain,
underestimates the true noise, and extrapolates a locally defined quantity to a
regime where it does not apply. A more careful but still inconsistent approach
measures the noise directly yet performs the measurement in the dark: the dark
noise is categorically different from the noise at the illumination level where the
responsivity is measured, and combining them yields a physically inconsistent
$D^{*}$ regardless of how carefully either quantity is measured individually. The
false signal-to-noise ratio that results from referencing an illuminated signal
peak to the dark noise floor is illustrated in Fig.~\ref{fig:PSD}: the discrepancy
grows with illumination as the noise floor rises with $I^{*}(\Gamma)$, and the
overestimate can be severe.

The only self-consistent protocol is to establish the operating point by applying
both the bias voltage and the DC illumination level at which the device is to be
characterised, and to measure noise and responsivity simultaneously at that point.
As illustrated in Fig.~\ref{fig:PSD}, this is achieved by modulating the optical
input at frequency $\omega_m$ with small amplitude $\delta\Gamma \ll \Gamma$ about
the DC level --- ensuring linear small-signal operation --- recording the PSD of
the output current, and extracting the tangential responsivity from the signal peak
and the local noise from the spectral floor in its vicinity. This yields a
$D^{*}_\mathrm{loc}$ that is internally consistent and directly comparable between
devices operating at the same point on the $I$-vs-$\Gamma$ curve.

\begin{figure}[t]
  \centering
  \includegraphics[width=\columnwidth]{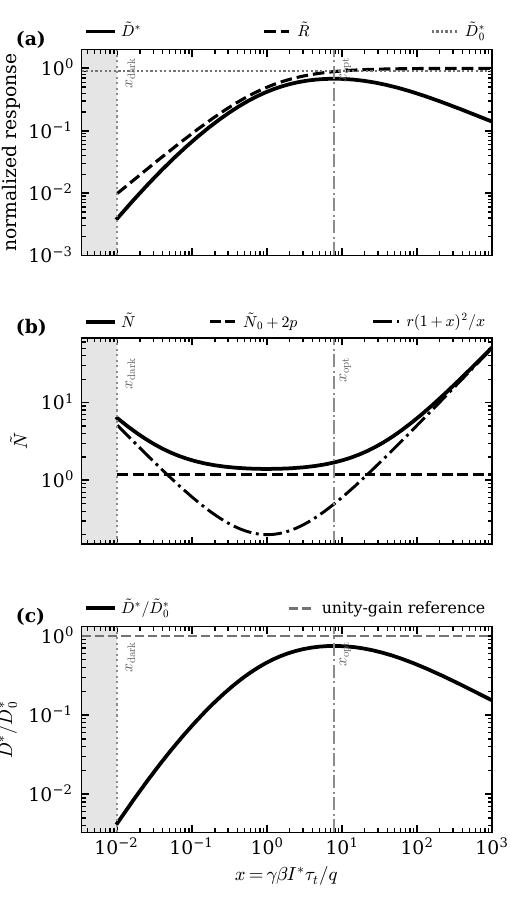}
  \caption{\textbf{Normalized responsivity, noise, and detectivity} versus the
    gain-state parameter $x = \gamma\beta I^{*}\tau_t/q$
    [Eqs.~(\ref{eq:Ntilde}),~(\ref{eq:Dtilde})]; $x$ parametrises the operating
    point, with a generally nonlinear relation to intensity (Sec.~III).
    (a)~$\tilde{R}$ saturates while $\tilde{D}^{*}$ peaks at $x^{*}$ and stays
    below the unity-gain reference $\tilde{D}^{*}_0 = 1/\sqrt{\tilde{N}_0 + 2p}$.
    (b)~The floor $\tilde{N}_0 + 2p$ is $x$-independent; the injection shot noise
    $r(1+x)^2/x$ is minimal at $x = 1$. (c)~$\tilde{D}^{*}/\tilde{D}^{*}_0<1$ at
    all operating points. The shaded region
    $x<x_\mathrm{dark}=\gamma\beta I_0\tau_t/q$ is inaccessible. Parameters:
    $\tilde{N}_0 = 1$, $p = 0.1$, $r = 0.05$.}
  \label{fig:detectivity}
\end{figure}

The appropriate sensitivity measure is therefore the local, small-signal
detectivity,
\begin{equation}
  D^{*}_\mathrm{loc}(\Gamma) = \frac{\sqrt{A}}{\mathrm{NEP}_\mathrm{loc}(\Gamma)},
  \label{eq:Dloc}
\end{equation}
with $\mathrm{NEP}_\mathrm{loc}$ given by Eq.~(\ref{eq:NEPtotal}). Using the
dimensionless variables $x$, $p$, $r$ and the normalization $\mathrm{NEP}_\star^2$,
the normalized local detectivity follows from Eq.~(\ref{eq:Ntilde}),
\begin{equation}
  \tilde{D}^{*}(x,p)
    = \frac{\tilde{R}(x)}{\sqrt{\tilde{N}(x,p)}}
    = \frac{x/(1+x)}{\sqrt{\tilde{N}_0 + 2p + r(1+x)^2/x}},
  \label{eq:Dtilde}
\end{equation}
where the normalized responsivity $\tilde{R}(x) = x/(1+x)$ is obtained by
normalizing $R(0)$ by its saturation value $R_\mathrm{max} = \eta q/\gamma h\nu$ as
$x \to \infty$. Equation~(\ref{eq:Dtilde}) is the central result of the detectivity
analysis and is illustrated in Fig.~\ref{fig:detectivity}.

The structure of Eq.~(\ref{eq:Dtilde}) encodes the competition between
amplification and noise. At small $x$ the responsivity grows linearly while the
injection shot noise $r/x$ diverges --- the injection current is too small to
transduce optical modulation efficiently. At large $x$ the responsivity saturates
while the injection shot noise grows as $rx$ --- the gain is exhausted and the
injection channel dominates the noise. The injection shot noise $r(1+x)^2/x$ has a
minimum at $x = 1$, and the detectivity consequently exhibits a finite maximum at
an optimal $x^{*}$. This confirms the observation made in Sec.~III that the
crossover $\gamma\beta I^{*}/q \sim 1/\tau_t$, corresponding to $x\sim 1$, is a
regime of particular significance: the optimal operating point lies in its
vicinity. Setting $d(\tilde{D}^{*2})/dx = 0$ yields the closed-form condition
\begin{equation}
  2(\tilde{N}_0 + 2p) = \frac{r(x^{*}-3)(x^{*}+1)^2}{x^{*}}.
  \label{eq:xopt}
\end{equation}
The right-hand side is positive only for $x^{*} > 3$, and since it increases
monotonically from zero at $x^{*} = 3$ to infinity as $x^{*}\to\infty$, there is
always exactly one solution $x^{*} > 3$ for any positive $\tilde{N}_0$, $p$, and
$r$. In physical terms, the optimal operating point always lies where
injection-assisted relaxation exceeds thermal emission by at least a factor of
three ($\gamma\beta I^{*}/q > 3/\tau_t$). Increasing $p$ or $\tilde{N}_0$ pushes
$x^{*}$ to larger values, while $r = \gamma/\beta$ sets the overall scale of the
injection noise penalty: a large $r$ means injected carriers are more likely to
relax the trap than to be collected, and the noise penalty per unit gain is high.

The optimization is subject to a physical lower bound: as $\Gamma\to 0$ the current
approaches $I_0$, giving $x_\mathrm{dark} = \gamma\beta I_0\tau_t/q$ as the minimum
accessible value of $x$. If $x^{*} < x_\mathrm{dark}$, the device cannot reach its
theoretical optimum and is injection-noise limited throughout. This identifies
minimizing $I_0$ through contact engineering as the primary route to improving
sensitivity in PM photodiodes.

We note that $\beta n_t$, the control parameter for linearity introduced in
Sec.~III, and $x = \gamma\beta I^{*}\tau_t/q$, the control parameter for gain
saturation and detectivity, are related but distinct: the former determines
whether the response is exponential, the latter whether the gain has saturated. A
device can be weakly coupled in the first sense yet sit at any value of the second,
and the two should not be conflated.

The finite optimum in Eq.~(\ref{eq:xopt}) already implies detectivity cannot grow
without bound with gain. The deeper result concerns the absolute ceiling. Consider
first an idealized noiseless, linear gain stage $G$ applied after the primary
detection event: $R_G = GR_0$, $S_G = G^2 S_0$, giving
\begin{equation*}
  \mathrm{NEP}^2_G = \frac{G^2 S_0}{G^2 R_0^2} = \mathrm{NEP}^2_0.
\end{equation*}
Even a perfectly noiseless gain stage cannot improve the intrinsic NEP. The reason
is fundamental: gain acts downstream of the primary photon-to-charge conversion and
cannot alter the statistics of the photoevents themselves. The quantity that
determines intrinsic sensitivity is the number of statistically independent
detection events registered per unit optical input --- set entirely by the quantum
efficiency $\eta$, not by any subsequent amplification. Increasing $\eta$ registers
more independent photoevents per incident photon, reducing the shot-noise
contribution to the NEP. Gain, operating on an already-formed electrical signal,
cannot recover information lost at the conversion step. In this thermodynamic
sense, quantum efficiency creates information at the point of measurement, whereas
gain merely redistributes what is already there.

Trap-assisted gain is subject to a stricter bound. It is not a noiseless amplifier
but a driven stochastic process maintained by electrical work from the applied
bias. The gain state $(n^{*}, I^{*})$ is a nonequilibrium dissipative state whose
maintenance requires continuous entropy production. The same injection process that
modulates the barrier and amplifies the current generates the injection shot noise
$r(1+x)^2/x$ --- a contribution strictly positive for any finite, physically
realizable set of device parameters, since $r = \gamma/\beta$ vanishes only in the
limits $\gamma\to 0$ or $\beta\to\infty$, neither of which corresponds to a device
that can reach a steady state or be fabricated. The input-referred NEP is therefore
\begin{equation}
  \mathrm{NEP}^2_\mathrm{gain} = \mathrm{NEP}^2_0 + \mathrm{NEP}^2_\mathrm{add},
  \qquad \mathrm{NEP}^2_\mathrm{add} \geq 0,
  \label{eq:NEPbound}
\end{equation}
with strict inequality for any physically realizable PM photodiode. It follows that
\begin{equation}
  D^{*}_\mathrm{gain} \leq D^{*}_\mathrm{intrinsic}.
  \label{eq:Dbound}
\end{equation}
This inequality is proven within the minimal single-trap model, but its
thermodynamic basis is general: any gain mechanism sustained by dissipative carrier
flux generates additional nonequilibrium fluctuations that cannot be engineered
away. As argued in Sec.~IV, the result is moreover insensitive to the energetic
disorder of the traps, which colours the noise spectrum but leaves the
input-referred bound intact. A concrete illustration follows. It has recently been
shown that the detectivity of organic and perovskite photodiodes is already several
orders of magnitude below the radiative limit, owing to Shockley--Read--Hall
recombination through mid-gap trap states~\cite{Sandberg2023}. Introducing a
trap-assisted gain mechanism to push the EQE beyond unity does not remedy this
deficit --- it compounds it. Gain, in this context, is not a path to recovering
lost sensitivity but a further departure from the radiative limit.

The result does not render gain without practical value. If readout electronics
contribute noise $S_\mathrm{ro}$ after the detector, the input-referred
contribution $S_\mathrm{ro}/G^2_\mathrm{eff}R_0^2$ decreases with increasing gain.
In a readout-noise-limited system, PM gain improves the system-level
signal-to-noise ratio by lifting the signal above the electronics noise floor. This
is its legitimate role --- not improving intrinsic photodetection sensitivity, but
relaxing requirements on subsequent electronics. Optimizing for maximum
$D^{*}_\mathrm{loc}$ at $x^{*}$ and optimizing for readout-limited performance are
different objectives, and conflating them produces devices that perform well on
reported figures of merit but poorly in the systems they are intended for.

Taken together, these results establish a coherent thermodynamic picture of gain in
PM photodiodes. The local detectivity, the optimal operating point, the
gain--bandwidth trade-off, and the absolute ceiling on sensitivity all emerge from
the same physics: the dissipative, nonequilibrium dynamics of a trap state driven
by optical generation and bias-sustained injection. The gain mechanism is not a
route to surpassing the fundamental limits of photodetection --- it is a driven
stochastic amplifier whose signal and noise are inseparably coupled through the
thermodynamics of the gain state. What it offers instead is a practical means of
suppressing readout noise, at the cost of nonlinearity, intensity-dependent
bandwidth, and an irreducible fluctuation penalty that grows with the gain itself.

\section{Conclusions}

We have employed a minimal analytical framework for trap-assisted
photomultiplication in photodiodes, adapted from the classical theory of barrier
photoconductivity, based on a single gain-active trapped-charge state that couples
photogenerated carriers to contact injection. The framework yields closed-form
expressions for the responsivity, noise, and local detectivity that are governed
throughout by the same trap-mediated dynamics, and makes no assumptions about
specific material systems or device architectures.

The gain mechanism is intrinsically self-limiting: the injection process that
amplifies the current simultaneously accelerates relaxation of the gain-enabling
state, saturating the effective gain and producing an inherently nonlinear,
operating-point-dependent response. The form of this nonlinearity is not universal.
Generalizing the single level to a distribution of traps and including bimolecular
recombination of the mobile carriers, the same mechanism produces superlinear,
linear, or strongly sublinear current--intensity characteristics. The sublinear
exponents are fixed by the energetic disorder ($T/T_c$) and the recombination order
(a factor of two between monomolecular and bimolecular control), while the barrier
coupling, permittivity, trap density, and Langevin reduction factor set only the
crossovers between regimes and whether the response ever turns superlinear.
Consequently the responsivity is not a device constant but a local, tangential
quantity that must be evaluated at a specified operating point; the chord gain is
not a meaningful or transferable figure of merit, and chord-gain comparisons across
the literature can be misleading. The linear dynamic range, as conventionally
defined, is either zero or confined to a narrow window, and should not be conflated
with the dynamic range.

The noise analysis reveals that internal gain introduces an irreducible fluctuation
penalty arising from the injection shot noise of the bias-sustained carrier flux.
The local detectivity consequently exhibits a finite optimum at $x^{*} > 3$, where
injection-assisted relaxation exceeds thermal emission by at least a factor of
three. The dark saturation current sets a physical lower bound on the accessible
operating range, identifying contact engineering to suppress dark injection as the
primary lever for improving sensitivity. 

The overarching thermodynamic result is that internal gain cannot improve the
intrinsic detectivity of a photodiode. Gain acts downstream of the primary
photon-to-charge conversion and cannot alter the statistics of that process; it
redistributes signal and noise within a driven dissipative system rather than
reducing the fundamental noise floor. Quantum efficiency, which acts at the
conversion step itself, is the quantity that genuinely controls intrinsic
sensitivity. These conclusions have direct implications for the interpretation of
reported $D^{*}$ values: a self-consistent measurement requires noise and
responsivity to be evaluated at the same illuminated operating point, and values
extracted using the shot-noise formula or dark-noise measurements at a different
operating point systematically overestimate the true sensitivity.

\FloatBarrier
\begin{acknowledgments}
The author is grateful to Johannes Benduhn for fruitful discussions and insightful feedback on this work.
\end{acknowledgments}

\bibliography{references}

\end{document}